\documentclass[11pt,reqno]{amsart}

\usepackage{graphicx}
\usepackage{amsmath,amstext,amsbsy,amssymb,color,caption}
\usepackage[cdot,amssymb]{SIunits}
\usepackage[latin1]{inputenc}
\usepackage{bm}
\DeclareMathOperator{\sign}{Sign} 

\title[]{\emph{The natural frequencies of
masonry beams}}

\author{MARIA GIRARDI}
\address{Istituto di Scienza e Tecnologie dell'Informazione "A. Faedo", CNR
Via G. Moruzzi 1\\
Pisa, 56124\\
Italy} \email{Maria.Girardi@isti.cnr.it}

\keywords{Nonlinear dynamics, slender masonry
structures, linear perturbation}

\def\thesubsection{\thesection\Alph{subsection}}
\numberwithin{equation}{section}
\def\theequation{\thesection\hbox{-}\arabic{equation}}

\begin{document} 

This is a preprint of an article published in \emph{Archive of
Applied Mechanics} (Springer). The final authenticated version is
available online at:
 https://doi.org/10.1007/s00419-021-01887-4

\begin{abstract}

The present paper aims at analytically evaluating  the natural
frequencies of cracked slender masonry elements.
The problem is dealt with in the framework of linear perturbation, and
the small oscillations of the structure are studied
under loaded conditions, after the equilibrium for permanent loads
has been achieved. A masonry beam element made of
no--tension (masonry--like) material is considered, and some explicit expressions of the beam's fundamental
frequency as a function of the external loads and the amplitude of
imposed deformations are derived. The analytical results are validated via
finite--element analysis.

\end{abstract}

\maketitle 

\section{Introduction}\label{sec:sec1}

The measurement of ambient vibrations has become a standard
procedure in Civil Engineering. In fact, these vibrations contain
precious information on both the structural behaviour and the health
status of buildings. Moreover, experimental frequencies and mode shapes can be
introduced in model updating procedures \cite{Friswell} and allow
estimating the mechanical properties and boundary conditions of such
structures, while  long--term measurements can help revealing the onset of
structural damage, through damage detection procedures.
In fact, modal properties are damage--sensitive features \cite{AGAR}, \cite{Z24}, \cite{SAL},

With regard to heritage masonry structures, the assumption of linear
elasticity, which usually underlies the study of their ambient
vibrations, may lead to errors. In fact, such structures are unable
to withstand large tensile stresses and are usually affected by
crack patterns.  These nonlinear effects have in general a
non--negligible influence on the structural stiffness and should not
be disregarded in the analysis. The dynamic behaviour of these structures
should be analyzed taking into account the existing cracks.

A great deal of effort has been devoted to numerically simulating the
effects of damage on structural vibrations. With regard to masonry
buildings, a common approach consists of simulating the cracks actually observed on the structure, by reducing the stiffness
of the elements of the finite--element model which  belong to the damaged parts \cite{PINEDA}, \cite{LOURENCO}. In \cite{bui} the dynamic properties of some masonry structures
at different damage levels are investigated via the discrete element
method. In \cite{Modal} a linear--perturbation numerical procedure is
presented, implemented in the NOSA--ITACA program \cite{SPRING}, to take into account the presence of cracks in the
calculation of a masonry structure's dynamic properties. The
procedure applies the constitutive equation of masonry--like materials
\cite{delpiero}, and the paper proves that the problem is governed by
the global tangent stiffness matrix, used in place of the linear
elastic one to evaluate the structure's modal properties. Some example applications are shown in
\cite{Mogadouro}, where linear perturbation is carried out to model
reinforcement operations on the Mogadouro tower in Portugal, and in 
\cite{FII}, where the
procedure is employed to reproduce the results of some laboratory tests conducted on a masonry arch
subjected to settlements of one support.

Similar problems also arise in the
study of the dynamic properties of  prestressed concrete cracked elements; the dependence of such structures' modal properties  on the external loads and  prestressing force level has long been debated, as shown in \cite{Abdel}, \cite{Hamed}, \cite{noble}.

In the present paper, an analytical approach is presented to evaluate the natural frequencies of masonry beams.
The small oscillations of such structures are studied
under loaded conditions, after the equilibrium for permanent loads
has been achieved. An explicit expressions of the beam's fundamental frequency as a
function of the external loads and  initial deformations is derived, by using a constitutive
relationship between the generalized deformation $\chi$ and the
generalized stresses $N$ and $M$ acting on the beam's sections.

The  simple case presented in the paper turns out to be of interest, since at the best of the Author's knowledge
the literature do not furnish yet explicit expressions  to estimate the effects of cracking on the natural frequencies of masonry structures.

The paper is organized as follows: in Section \ref{sec:sec2} the problem is introduced.
Section \ref{sec:sec3} briefly recalls the constitutive equation of masonry--like beams and in Section \ref{sec:sec4} some example applications are presented. Finally, Section \ref{sec:sec5} is devoted to testing of the analytical results via finite--element analysis.

\section{Small oscillations of a loaded masonry beam}
\label{sec:sec2}

Let us consider a rectilinear beam with length $l$ and rectangular
cross section with height $h$ and width $b$, subjected along its
axis to a constant axial force $N$.  Let us denote by $E$ and $\rho$
the Young's modulus and the density of the material, respectively,
and by $J = bh^3/12$ the moment of inertia of the beam's cross
section. Let $x$ be the abscissa along the beam's axis and $t$ the
time.

The curvature of the beam is denoted by $\chi$ and, under the
Euler--Bernoulli hypothesis, the bending moment $M(\chi):R\to R$ is
a continuous differentiable function, whose second derivative is
assumed to be piecewise continuous.

We assume the effects of both the shear strain and the rotary
inertia to be negligible. Moreover, we limit ourselves to
considering situations in which the transverse displacement
$\phi(x,t)$ and its derivative $\phi(x,t)_x$ are small, so that we
can neglect the effects of the axial force on the dynamic
equilibrium of the beam and write
\begin{equation}
\chi(x,t)=-\phi_{x,x}(x,t)\label{eq:eq1}.
\end{equation}

If we introduce function
\begin{equation}
f(\chi)=\frac{M(\chi)}{\rho\,b\,h},\label{eq:eq2}
\end{equation}
the motion of the beam is expressed by

\begin{equation}
\phi_{t,t}-(f(\chi))_{x,x}=q(x,t)\label{eq:eq3},
\end{equation}
where
\begin{equation}
q(x,t) = \frac {p(x,t)}{\rho\,b\,h}
\end{equation}

and $p(x,t)$ is the transverse load per unit length.

Let us consider, at $t=0$, a load $\bar q(x)$ inducing in the beam
an initial deformation $\bar \phi(x)$ and curvature change
$\bar\chi(x)$. The beam reaches equilibrium under the load $\bar q$,
and thus

\begin{equation}
-f(\bar\chi)_{x,x}=\bar q(x)\label{eq:eq3b}.
\end{equation}

We are interested in studying the small oscillations $\delta\phi$ of
the beam around $\bar\phi$.




\begin{equation}
\delta\phi_{t,t}-\left(\frac{\partial f}{\partial\chi}
\Bigg\vert_{\bar\chi}\delta\chi\right)_{x,x}=0 \label{eq:eq6},
\end{equation}
where we use the approximation

\begin{equation}
f(\bar\chi+\delta\chi)\simeq f(\bar\chi)+\frac{\partial
f}{\partial\chi} \Bigg\vert_{\bar\chi}\delta\chi. \label{eq:eq5}
\end{equation}

In case of a simply supported beam, since we limit ourselves to
studying the fundamental frequency of the beam, we can approximate
the small oscillations $\delta\phi$ as follows:

\begin{equation}
\delta\phi\simeq \delta a \sin\left(\frac{\pi}{L} x\right) u(t),
\label{eq:eq7}
\end{equation}
where $\delta a>0$ represents the amplitude of the beam's
oscillation around $\bar\phi$. In view of \eqref{eq:eq7},
\eqref{eq:eq6} becomes

\begin{equation}
\sin\left(\frac{\pi}{L} x\right)\ddot
 u-\left(\frac{\pi}{L}\right)^2\left(\frac{\partial f}{\partial\chi}
\Bigg\vert_{\bar\chi} \sin{\left(\frac{\pi}{L}x\right)}\right)_{x,x}
u=0 \label{eq:eq8}.
\end{equation}

Let us  now use the Galerkin method, multiply \eqref{eq:eq8} by
$\sin{\left(\frac{\pi}{L}x\right)}$ and integrate over the beam's
length. The first member of the equation becomes:

\begin{eqnarray*}
&\ddot u\int_{0}^{L}\sin\left(\frac{\pi}{L} x\right)^2
dx-\frac{\pi^2}{L^2}\left(\int_{0}^{L}\sin\left(\frac{\pi}{L}
x\right)\left(\frac{\partial f}{\partial\chi} \Bigg\vert_{\bar\chi}
\sin{\left(\frac{\pi}{L}x\right)}\right)_{x,x} dx\right) u =\\
&= \frac{L}{2} \ddot u-
\frac{\pi^2}{L^2}\left(\sin{\left(\frac{\pi}{L}x\right)\left(\frac{\partial
f}{\partial\chi} \Big\vert_{\bar\chi}
\sin{\left(\frac{\pi}{L}x\right)}\right)_{x}}\Bigg\vert_{0}^{L}-\int_{0}^{L}\frac{\pi}{L}\cos\left(\frac{\pi}{L}
x\right)\left(\frac{\partial f}{\partial\chi} \Bigg\vert_{\bar\chi}
\sin{\left(\frac{\pi}{L}x\right)}\right)_{x}
dx\right)u=\\
&= \frac{L}{2} \ddot u+
\frac{\pi^3}{L^3}\left(\int_{0}^{L}\cos\left(\frac{\pi}{L}
x\right)\left(\frac{\partial f}{\partial\chi} \Bigg\vert_{\bar\chi}
\sin{\left(\frac{\pi}{L}x\right)}\right)_{x} dx\right)u=\\
&=\frac{L}{2} \ddot u+ \frac{\pi^3}{L^3}\left(
\cos{\left(\frac{\pi}{L}x\right)\frac{\partial f}{\partial\chi}
\Big\vert_{\bar\chi}
\sin{\left(\frac{\pi}{L}x\right)}}\Bigg\vert_{0}^{L}+\frac{\pi}{L}\int_{0}^{L}\sin^2\left(\frac{\pi}{L}
x\right)\frac{\partial f}{\partial\chi} \Bigg\vert_{\bar\chi}
dx\right)u,
\end{eqnarray*}

where we used the boundary conditions
$\sin{\left(\frac{\pi}{L}x\right)}\Bigg\vert_{0}^{L}=0$. Thus,
equation \eqref{eq:eq8} becomes

\begin{equation}
\ddot u+
\frac{2\pi^4}{L^5}\left(\int_{0}^{L}\sin^2\left(\frac{\pi}{L}
x\right)\frac{\partial f}{\partial\chi} \Bigg\vert_{\bar\chi}
dx\right)u=0. \label{eq:eq9}
\end{equation}
 The fundamental frequency can be directly obtained  from
 equation \eqref{eq:eq9}, once the constitutive equation $f(\chi)$
 is given.

 \begin{figure}
\centering
\includegraphics[width=11cm]{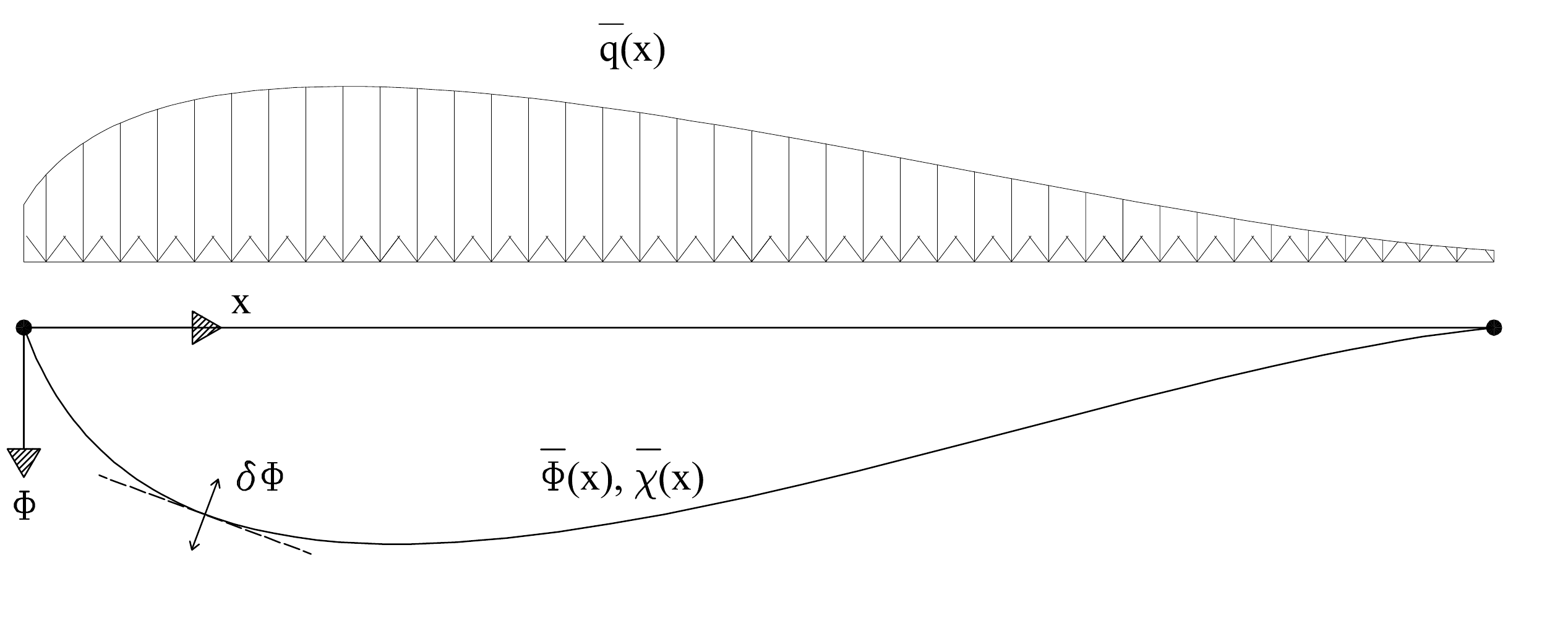}
\caption{The beam model.}\label{fig:fig8}
\end{figure}

\section{A constitutive equation for masonry--like beams}
\label{sec:sec3}

In this section we briefly recall a simple constitutive equation for
one--dimensional masonry--like elements with rectangular cross
section. Despite its simplicity, this constitutive equation
effectively describes the mechanical behaviour of a masonry beam
subjected to axial force and bending moment. The beam's cross
section is fully compressed since the vertical load falls inside the
central nucleus of inertia, and reaches the limit elastic behaviour
when the vertical load acts on the perimeter. Many authors have made
use of this simplified equation to find analytical and numerical
solutions to the static and the dynamic behaviour of masonry beams
and arches (some examples can be found in
\cite{Barsotti},\cite{ADF},
\cite{DeFalco2007},\cite{Pintucchi},\cite{Zani}). In
\cite{GirardiandLucchesi} and \cite{Girardi2014} some explicit
approximate solutions are proposed for modelling the free and forced
oscillations of masonry beams.

Under the classical hypothesis of Euler--Bernoulli and providing
that the axial force $N$ along the beam's axis is known, infinite
compressive strength is allowed, and the material is unable to
withstand tensile stresses, function $f(\chi)$ becomes:

\begin{equation}
\label{eq:eq10} f(\chi)=
\begin{cases}
\quad c^2 \chi \quad &\text{for $\left| \chi \right|
\le \alpha$},\\
\quad  c^2 \alpha   \sign(\chi )  (3 - 2\sqrt {\frac{\alpha
}{{\left| \chi  \right|}}} )\quad &\text{for} \left| \chi \right|
> \alpha,
\end{cases}
\end{equation}
where

\begin{equation}\label{eq:eq11}
\alpha=-\frac{2N}{Ebh^2}
\end{equation}
is the curvature corresponding to the elastic limit and $c^2=
\frac{EJ}{\rho b h}$  is the elastic constant of the beam.

 Function \eqref{eq:eq10} is
plotted in Figure \ref{fig:fig8}. In the nonlinear region, for
$\left| \chi \right|
> \alpha$, increasing values of $\chi$  correspond to a rapid
decrease  in the section's stiffness, and the bending moment tends
toward its limit value $\left|Nh/2\right|$. Moreover, $M(\chi)$ is
continuous with its first derivative, while the second derivative
undergoes a jump at $\left|\chi\right|=\alpha$.

\begin{figure}
\centering
\includegraphics[width=15cm]{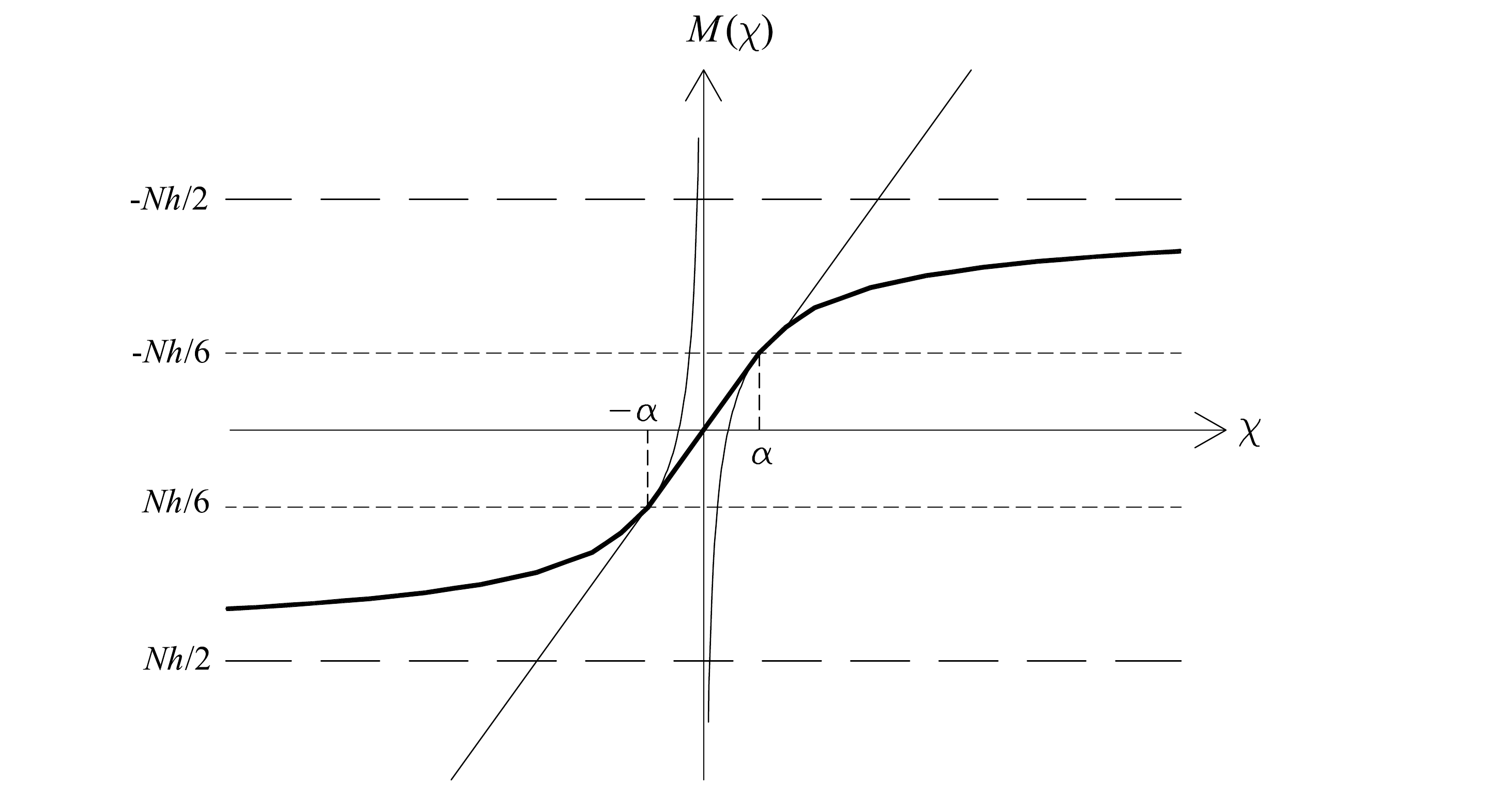}
\caption{The constitutive equation $M-\chi$ for a rectangular
section with infinite compressive strength and zero tensile
stress.}\label{fig:fig8}
\end{figure}

Equation \eqref{eq:eq10} is used in the following in conjunction
with \eqref{eq:eq9} to explicitly evaluate the fundamental frequency
of masonry beams subjected to different loading conditions. The
first derivative of $f$ with respect to $\chi$, which appears in
\eqref{eq:eq9}, is:

\begin{equation}
\label{eq:eq12} f'(\chi)=
\begin{cases}
\quad c^2  \quad &\text{for $\left| \chi \right|
\le \alpha$},\\
\quad  c^2 \sqrt{\frac{\alpha^3}{{\left| \chi \right|}^3}}\quad
&\text{for} \left| \chi \right|
> \alpha.
\end{cases}
\end{equation}

%

\section{Example applications}
\label{sec:sec4}

\subsection{Case 1: constant curvature}
\label{subsec:subsec4-1}

Let us consider the case of a masonry beam in an initial deformed
shape with constant curvature $\bar\chi$. This is the simple, but
meaningful case of a beam subjected to a  vertical load acting with
constant eccentricity $e$  along the axis. In this case, the bending
moment acting on the beam assumes the constant value

\begin{equation}
\label{eq:eq13} M= N\cdot e,
\end{equation}

\noindent and nonlinear behaviour of the structure is attained for
the eccentricity

\begin{equation}
\label{eq:eq14} \left| e\right| = \frac{h}{6}.
\end{equation}

Thus, from \eqref{eq:eq9} and \eqref{eq:eq12} the fundamental
frequency $\omega$ of the beam (in radians per second) is:

\begin{eqnarray}
\label{eq:eq15} \omega^2= \frac{\pi^4}{L^4}\,c^2 =
\omega_{el}^2\quad &\text{for} \left| \bar\chi \right|
\le \alpha,\\
\label{eq:eq15b} \omega^2=  \frac{2\pi^4 c^2}{L^5}\int_0^L
\sin^2\left(\frac{\pi x}{
L}\right)\sqrt{\frac{\alpha^3}{\left|\bar\chi\right|^3}}\,dx\quad
&\text{for} \left| \bar\chi \right|
> \alpha,
\end{eqnarray}

\noindent and $\bar\chi$, given the bending moment \eqref{eq:eq13} ,
can be deduced  from \eqref{eq:eq10}
\begin{equation}
\label{eq:eq16} \left| \bar\chi\right|=\frac{4 \,\alpha^3
\,c^4}{(\left| f\right|-3\,\alpha \,c^2)^2}.
\end{equation}

Equation \eqref{eq:eq15} coincides with the fundamental frequency of
a linear elastic beam \cite{Clough} and can be deduced by means of
\eqref{eq:eq9}, per $\chi=\alpha$ . When the eccentricity reaches
the border of the section,  $\left| e\right| = h/2$, the frequency
of the beam tends to zero.

It is worth noting that the value of the linearized frequency
\eqref{eq:eq15b} does not depend on the normal force $N$, but only
on the eccentricity $e$. In fact, using \eqref{eq:eq13} and
\eqref{eq:eq16}, after simple calculations, the frequency assumes
the following expression:

\begin{eqnarray}
\label{eq:eq17} \omega= \frac{\pi^2}{L^2}\,c=\omega_{el} \quad
&\text{for} \left| e \right|
\le \frac{1}{6},\\
\label{eq:eq17a}
\omega=\frac{3}{4}\,\frac{\pi^2c}{L^2}\sqrt{6\left(1-2\frac{\left|
e\right|}{h}\right)^3}
=\frac{3}{4}\,\omega_{el}\sqrt{6\left(1-2\frac{\left|
e\right|}{h}\right)^3} \quad &\text{for} \frac{1}{6}<\left| e
\right| \le \frac{1}{2}.
\end{eqnarray}

\begin{figure}
\centering
\includegraphics[width=11cm]{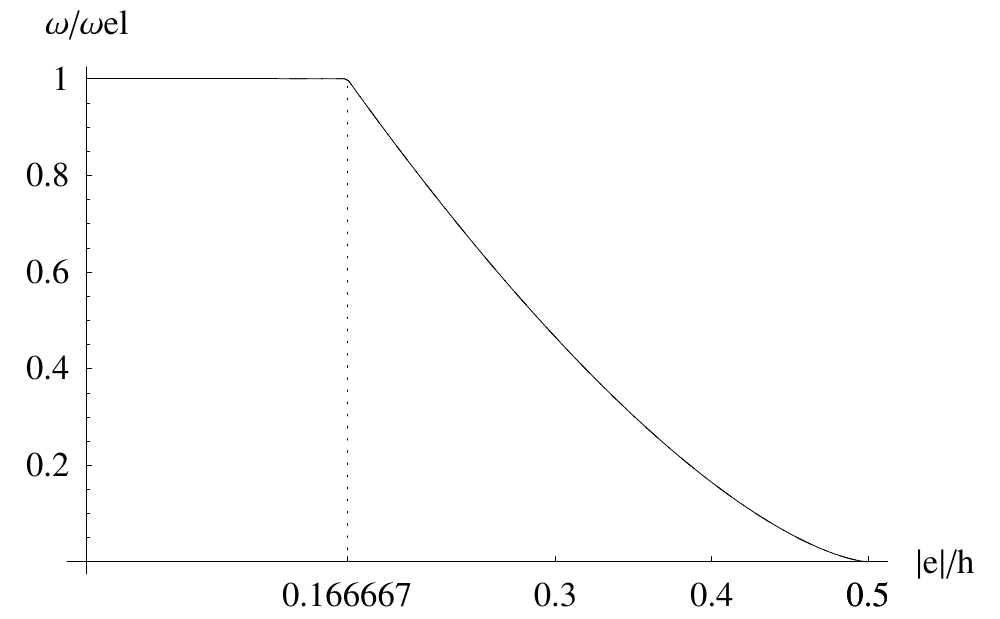}
\caption{Case 1: ratio $\omega el/\omega$ vs. eccentricity $|e|$ of
the axial load $N$. }\label{fig:figcase1}
\end{figure}

Equations \eqref{eq:eq17} and \eqref{eq:eq17a} are plotted in Figure
\ref{fig:figcase1}.

\subsection{Case 2: uniform transverse load}
\label{subsec:subsec4-2}

Let us consider the case of a beam subjected to a uniform transverse
load $p$  and axial force $N$.

 \begin{figure}[b]
\centering
\includegraphics[]{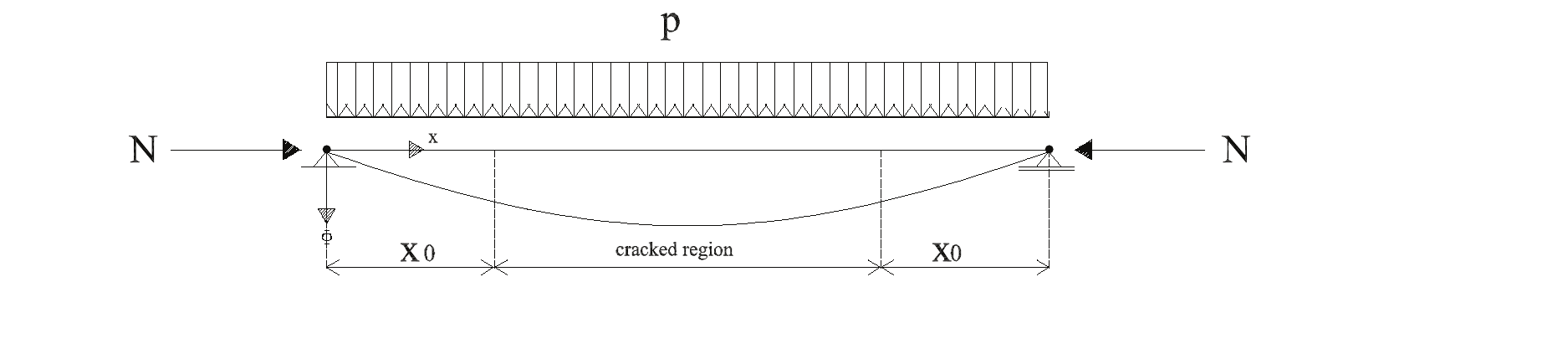}
\caption{Case 2: uniform transverse load.}\label{fig:fig8b}
\end{figure}

The beam enters the nonlinear field when $p$ reaches the value

\begin{equation}
\label{eq:eq17c} \bar p=\frac{4\left| N\right|\,h}{3\,L^2}.
\end{equation}

For greater values of the transverse load, the beam's axis can be
divided into three regions: the central part of the beam  is in the
nonlinear field, while two lateral regions, symmetric with respect
to the midsection, still exhibit linear behaviour (see Figure
\ref{fig:fig8b}). Let us denote by $x_0$ the abscissa of the beam's
section with curvature $\chi=\alpha$, $0\le x_0\le L/2$. Function
$x_0$ can be easily deduced:

\begin{equation}
\label{eq:eq18} x_0=
\begin{cases}
\quad \frac{L}{2} \quad &\text{for $p < \bar p$},\\
\quad \frac{L}{2}-\frac{L}{2}\sqrt{1-\frac{\bar p}{p}} \quad
&\text{for $p\ge\bar p$}.
\end{cases}
\end{equation}

\begin{figure}
\centering
\includegraphics[width=11cm]{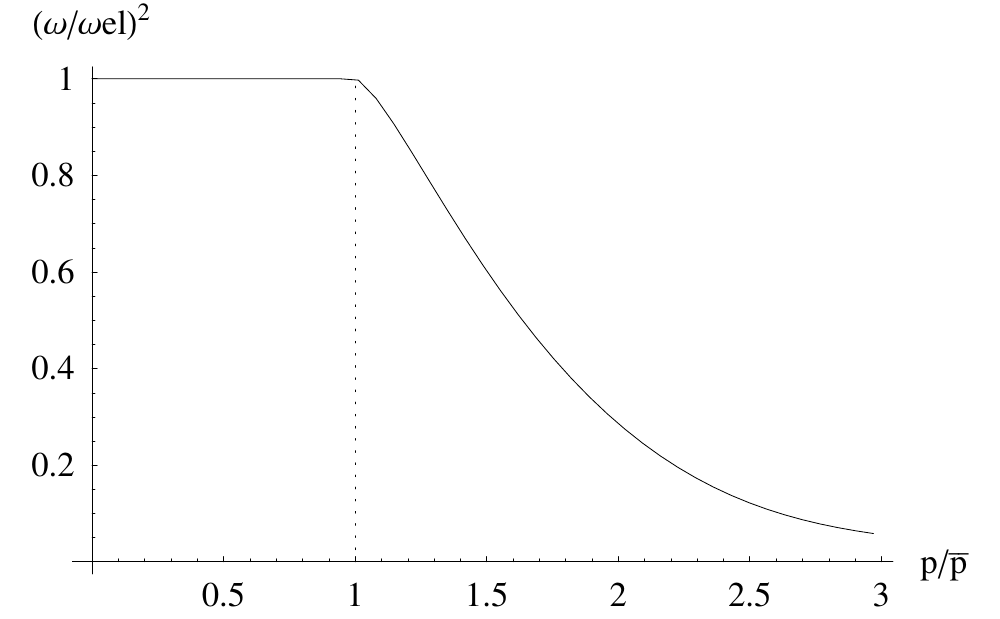}
\caption{Case 2: ratio $(\omega el/\omega)^2$ vs. ratio $p/\bar p$.
}\label{fig:figcase2}
\end{figure}

In view of \eqref{eq:eq9}, the expression for the fundamental
frequency becomes, for $p\ge\bar p$

\begin{equation}
\label{eq:eq19} \omega^2=  \frac{4\pi^4 c^2}{L^5}\left(\int_0^{x_0}
\sin^2\left(\frac{\pi x}{
L}\right)\,dx+\int_{x_0}^{L/2}\sin^2\left(\frac{\pi x}{
L}\right)\sqrt{\frac{\alpha^3}{\left|
\bar\chi\right|^3}}\,dx\right),
\end{equation}

\noindent with $\left| \bar\chi\right|$ given by \eqref{eq:eq16}.
After simple calculations, equation \eqref{eq:eq19} can be expressed
in the following form

\begin{equation}
\label{eq:eq19ad} \omega^2 = 4 \omega_{el}^2\left(\int_0^{x_0/L}
\sin^2(\pi\, y)\, dy + 8\int_{x_0/L}^{1/2}\sin^2(\pi\,
y)\left|\frac{p}{\bar p}(y-y^2)-\frac{3}{4}\right|^3\,dy \right),
\end{equation}

\noindent with $y = x/L$. Thus, ratio $\omega/\omega_{el}$, shown in
Figure \ref{fig:figcase2}, turns out to be dependent only on the
ratio $p/\bar p$.

The mid--section of the beam reaches the limit bending moment for
the transverse load

\begin{equation}
\label{eq:eq20} \bar{\bar p}=\frac{4\left| N\right|\,h}{L^2}=3\,\bar
p.
\end{equation}

Accordingly, abscissa $x_0$ attains the value

\begin{equation}
\label{eq:eq21} x_0=\frac{L}{2}-\frac{L}{\sqrt 6},
\end{equation}

\noindent which is independent of the transverse load and the normal
force acting and corresponds to about one tenth of the beam's total
length. In this case, equation \eqref{eq:eq19ad} reduces to

\begin{equation}
\label{eq:eq22} \left(\omega/\omega_{el}\right)^2 = 4
\left(\int_0^{\frac{1}{2}-\frac{1}{\sqrt{6}}} \sin^2(\pi y)\, dy +
8\int_{\frac{1}{2}-\frac{1}{\sqrt{6}}}^{1/2}\sin^2(\pi
y)\left|3\,(y-y^2)-\frac{3}{4}\right|^3\,dy \right)\simeq 0.05.
\end{equation}


\subsection{Case 3: Initial deformed shape}
\label{subsec:subsec4-3}

In this last case, let us impose on the beam an initial deformation
defined by the following equation:

\begin{equation}
\label{eq:eq22} \bar \phi(x)=A  \,\sin\left(\frac{\pi x}{L}\right)
\end{equation}

\noindent with $A\ge 0$. Accordingly,  curvature $\bar\chi$ assumes
the value

\begin{equation}
\label{eq:eq23} \bar \chi(x)= \frac{A\pi^2}{L^2} \sin\left(\frac{\pi
x}{L}\right),
\end{equation}
and, setting $\bar\chi=\alpha$, abscissa  $x_0$ becomes

\begin{equation}
\label{eq:eq23}x_0= \frac{L}{\pi}  \arcsin\left(\frac{\alpha
L^2}{A\pi^2}\right).
\end{equation}

\noindent Function $x_0$ is $L/2$ for

\begin{equation}\label{eq:eq24}
A_m=\alpha L^2/\pi^2,
\end{equation}

\noindent and tends to zero for $A\to\infty$.

\begin{figure}
\centering
\includegraphics[width=11cm]{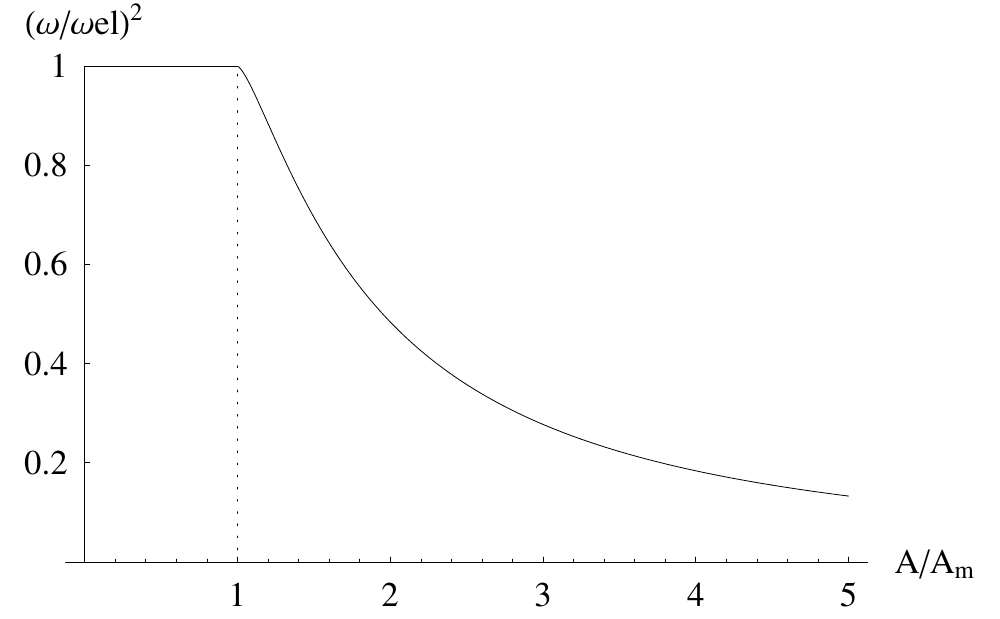}
\caption{Ratio $(\omega el/\omega)^2$ vs. ratio $A/A_m$.
}\label{spostin}
\end{figure}

In this case, equations \eqref{eq:eq9} and \eqref{eq:eq12} yield:

\begin{equation} \label{eq:eq25} \begin{split}
\omega^2=&  \frac{2\pi^4 c^2}{L^5}\left(2\int_0^{x_0}
\sin^2\left(\frac{\pi x}{
L}\right)\,dx+\int_{x_0}^{L-x_0}\sin^2\left(\frac{\pi x}{
L}\right)\sqrt{\frac{\alpha^3 L^6}{A^3\pi^6
\sin\left(\frac{\pi x}{L}\right)^3}}\,dx\right)=\\
=&\frac{2\pi^4
c^2}{L^5}\left(x_0-\frac{L^3\alpha}{A\pi^3}\sqrt{1-\frac{L^4\alpha^2}{A^2\pi^4}}+\frac{L^3\alpha}{A\pi^3}\int_{x_0}^{L-x_0}\sqrt{\frac{\alpha}{A}\sin\left(\frac{\pi
x}{L}\right)}\,dx\right)=\\
=&2\,\omega_{el}^2\left(\frac{x_0}{L}-\frac{A_m}{\pi
A}\sqrt{1-\left(\frac{A_m}{A}\right)^2}+\int_{\frac{x_0}{L}}^{1-\frac{x_0}{L}}\sqrt{\left(\frac{A_m}{A}\right)^3\sin(\pi
y)}\,dy\right),
\end{split}
\end{equation}
where the first two terms regard the unfractured part of the beam,
while the third is for the fractured part.  For $A=A_m$ frequency
\eqref{eq:eq25} reduces to the linear elastic value $\omega_{el}$,
while for $A\to\infty$ the frequency tends to zero. Function $x_0/L$
depends solely on the variable $A_m/A$. Thus, as for the previous
cases, expression \eqref{eq:eq25} can be expressed in the
non--dimensional form shown in Figure \ref{spostin}.

\section{Numerical solutions}
\label{sec:sec5}

The analytical solutions have been tested against finite element
analysis. To this end, we adopt a conforming beam element equipped
with cubic polynomial shape functions (Hermite polynomials), thereby
guaranteeing continuity of both the transverse displacements and
rotations of the beam's axis \cite{OAC}. Accordingly, the beam
element has two degrees of freedom for each node: one for the beam's
transverse displacement $\phi$, the other associated to its
derivative $\phi_x$.

The first application of this element to masonry--like beams is
reported on in
 \cite{Pintucchi}, \cite{PintucchiPhD}, where the element is
described in detail. Here the main ingredients of the calculation
are briefly recalled. The core of the analysis is calculation of the
element tangent stiffness matrix, which for constitutive equation
\eqref{eq:eq10} takes the explicit form


\begin{equation}
\label{eq:eq26} \mathbf{K}^e_T=  \frac{EJ}{c^2}\int_e
\bm{\Psi}''\bm\Psi''^{\text{T}}\,f'\,d\xi
\end{equation}

\noindent where 
$\bm\Psi$ is the vector of the shape functions, $f'(\chi)$ is given
by equations \eqref{eq:eq12}, and relation \eqref{eq:eq1} holds
true. The vector of the element's internal forces  is


\begin{equation}
\label{eq:eq27} \mathbf{f}_i^{\,e}=  \frac{EJ}{c^2} \int_e
\bm\Psi''\,f\,d\xi,
\end{equation}
where $f$ is given by  \eqref{eq:eq10}, while the vector of the
external loads acting on the element and the mass matrix of the
element assume the usual expressions

\begin{equation}
\label{eq:eq28} \mathbf{f}_e^{\,e}= \int_e \bm\Psi p\, d\xi,
\end{equation}

\begin{equation}
\label{eq:eq29} \mathbf{M}^e=  \rho b h \int_e
\bm{\Psi}\,\bm{\Psi}^T d\xi.
\end{equation}

The numerical procedure adopted here follows the method shown in
\cite{Modal} and is known as linear perturbation. This method
consists of the following steps:

\textbf{Step 1.} Model the structure via finite elements.

\textbf{Step 2.} Apply external loads and solve the nonlinear
equilibrium problem through an iterative scheme. Extract the tangent
stiffness matrix calculated in the last iteration before
convergence.

\textbf{Step 3.} Perform a modal analysis about the equilibrium
solution,  using the tangent stiffness matrix in place of the linear
elastic one.

\vspace{1\baselineskip}

 Herein the finite element \eqref{eq:eq26}--\eqref
{eq:eq29}, implemented in the Mathematica environment, is used to
model the masonry beam. For each load case, the tangent stiffness
matrix is calculated via a Newton--Raphson scheme, and the
corresponding value of the fundamental frequency is evaluated. This
value is then compared to the analytical solutions reported in
Section \ref{sec:sec4}.

The numerical procedure is applied to a masonry--like beam--column
with the following geometric characteristics:

\begin{equation} \label{eq:eq30}
\begin{split}
&L=\unit{6}{\metre},\\
&h= \unit{0.4}{\metre},\\
&b=\unit{1}{\metre},\\
&\rho=\unit{1800}{\kilogram\per\metre^3},\\
&E=\unit{3\cdot 10^9}{\pascal},\\
\end{split}
\end{equation}

The beam is simply supported at the ends and discretized into $30$
elements (different numbers of elements have been also tested to
check software performance). The example applications presented
regard the three cases presented in Section \ref{sec:sec4}.

\subsection{Case 1: constant curvature}
\label{subsec:subsec5-1}

Beam \eqref{eq:eq30} is subjected to increasing values of the
eccentricity. The normal force acting on the model is
$N=\unit{-500000}{\newton}$.  Different tests, not reported on here,
have been performed to check the influence of the normal force on
the beam's fundamental frequency given eccentricity $e$: the
numerical value of the frequency turned out to be  independent of
the normal force, as predicted by the analytical solution.

\begin{figure}
\centering
\includegraphics[width=11cm]{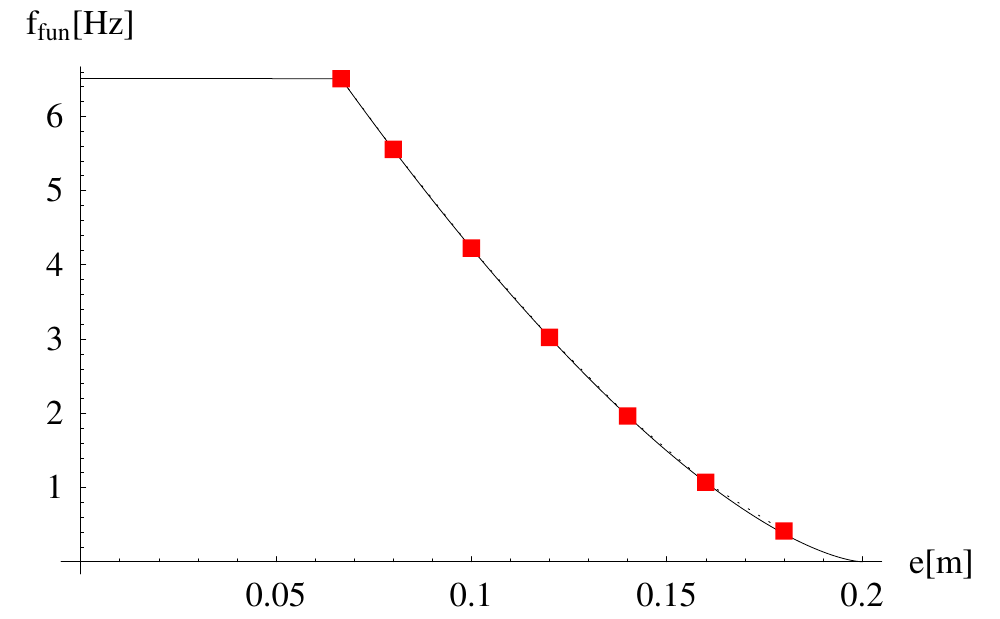}
\caption{Fundamental frequency $\text{f}_{\text{fun}}$ vs. the
eccentricity $e$ of the normal force. Analytical solution in
continuous line, while boxes represent the FE
simulation.}\label{mcostvar_num}
\end{figure}

Figure \ref{mcostvar_num} shows the beam's fundamental frequency
$f_{fun}$ versus the eccentricity $e$ of the normal force. The
analytical solution \eqref{eq:eq17a}, \eqref{eq:eq17} is plotted as
a continuous line, while boxes are for the finite element
simulation.


\subsection{Case 2: uniform transverse load}
\begin{figure}
\centering
\includegraphics[width=11cm]{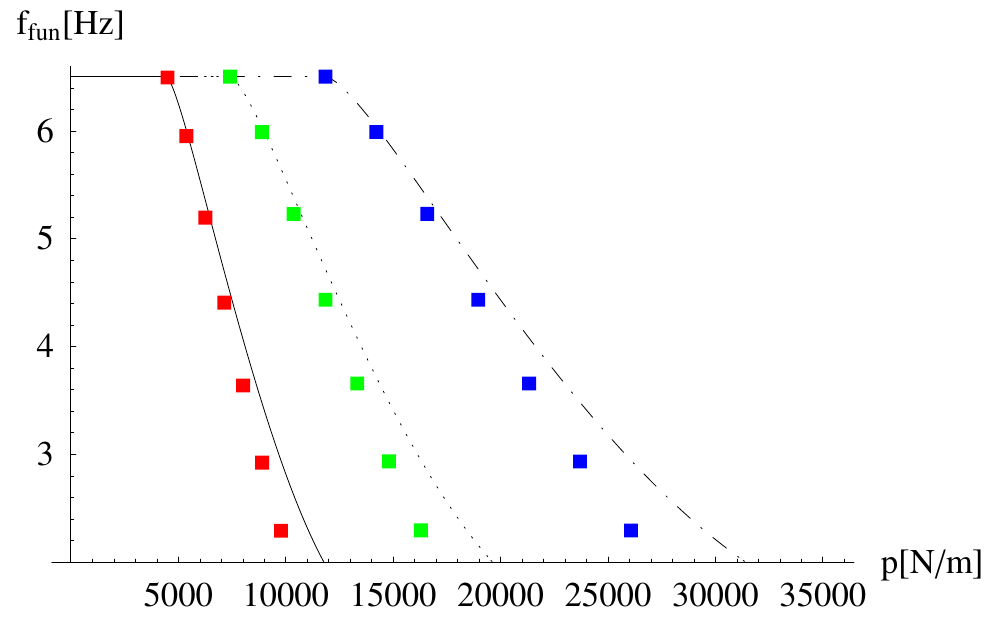}
\caption{Fundamental frequency $\text{f}_{\text{fun}}$ vs.
transverse load $p$. Continuous line: $N=\unit{-300000}{\newton}$.
Dotted line: $N=\unit{-500000}{\newton}$. Dash--dotted:
$N=\unit{-800000}{\newton}$. Boxes represent the FE
simulation.}\label{pvar_num}
\end{figure}

Beam \eqref{eq:eq30} is subjected to a transverse load $p$ of
increasing amplitude, with a normal force $N$ acting along the beam.
Different values of $N$  are tested. The beam's fundamental
frequency is shown in Figure \ref{pvar_num}, where lines represent
the analytical solution \eqref{eq:eq19}, while boxes represent the
results of the finite element analysis.

\label{subsec:subsec5-2}

\subsection{Case 3: initial deformed shape}
\label{subsec:subsec5-3} 
\begin{figure}
\centering
\includegraphics[width=11cm]{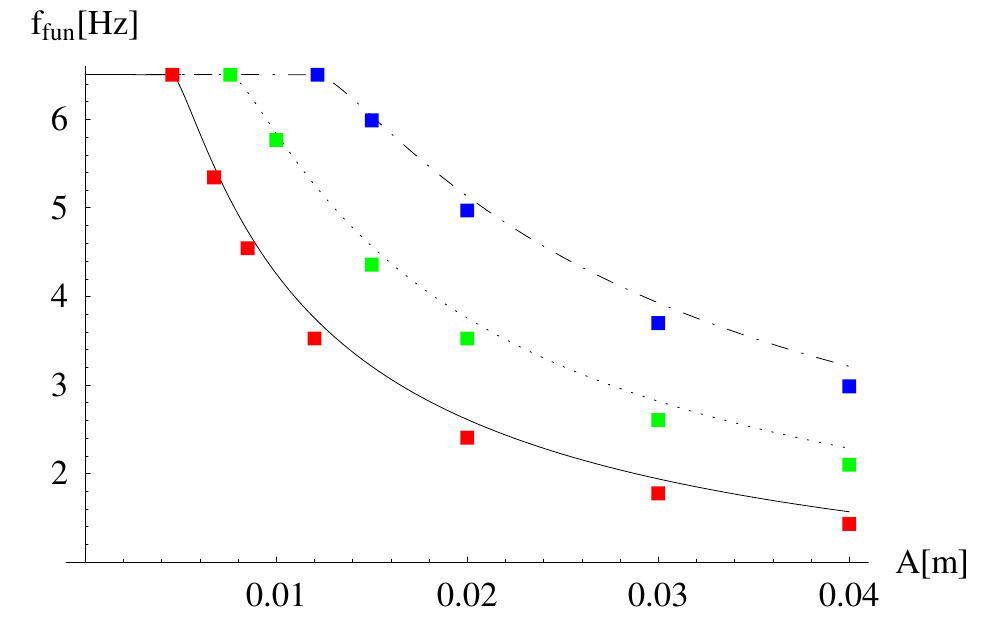}
\caption{Fundamental frequency $\text{f}_{\text{fun}}$ vs. the
amplitude $A$ of the initial displacement. Continuous line:
$N=\unit{-300000}{\newton}$. Dotted line:
$N=\unit{-500000}{\newton}$. Dash--dotted:
$N=\unit{-800000}{\newton}$. Boxes represent the FE
simulation.}\label{amplin_num}
\end{figure}

Beam \eqref{eq:eq30} is in an initial
deformed shape defined by equation \eqref{eq:eq22}. Increasing
values of the maximum amplitude $A$ are evaluated for different
values of the normal force $N$ acting along the beam. The beam's
fundamental frequency is shown in Figure \ref{amplin_num}, where
lines represent the analytical solution \eqref{eq:eq25}, while boxes
represent the results of the finite element analyses.

\section*{Conclusions}\label{sec:sec6}

An analytical approach to determine the fundamental frequency of
masonry beam--columns in the presence of cracks has been presented.
The study has been carried out under the hypothesis of masonry--like
material. Some explicit expressions have been obtained, with the
frequency depending on the external loads acting or the initial
deformation imposed on the beam. The analytical method has been
validated via finite--element analysis, and such a comparison
reveals the good agreement found between between the numerical and
analytical results.

\end{document}